\documentclass[prd,preprint,12pt]{article}
\usepackage[dvips]{graphicx}
\usepackage{endnotes}
\usepackage{amssymb}
\usepackage{amsmath}
\usepackage{graphicx}
\usepackage{multicol}

\begin{document}
\title{Observational constraints on the interacting Ricci dark energy model}
\author{Masashi Suwa$^{1}$\footnote{email: suwa.masashi@nihon-u.ac.jp}  and Takeshi Nihei$^{2}$\footnote{email: nihei@phys.cst.nihon-u.ac.jp}}
\date{\it \small $^1$College of Pharmacy, Nihon University, 7-7-1, Narashinodai, Funabashi-shi, Chiba, 274-8555, Japan\\
$^2$Physics Department, College of Science and Technology, Nihon University, 
1-8-14, Kanda-Surugadai, Chiyoda-ku, Tokyo 101-8308, Japan}
\maketitle
\begin{abstract}
We consider an extension of the holographic Ricci dark energy model by introducing an interaction between dark energy and matter. In this model, the dark energy density is given by $\rho_{\Lambda}=-\frac{1}{2}\alpha M_{p}^{2}{\cal R}$, where ${\cal R}$ is the Ricci scalar curvature, $M_{p}$ is the reduced Planck mass, and $\alpha$ is a dimensionless parameter. The interaction rate is given by $Q=\gamma H \rho_{\Lambda}$, where $H$ is the Hubble expansion rate, and $\gamma$ is a dimensionless parameter. We investigate current observational constraints on this model by applying the type Ia supernovae, the baryon acoustic oscillation and the CMB anisotropy data. It is shown that a nonvanishing interaction rate is favored by the observations. The best fit values are $\alpha=0.45\pm 0.03$ and $\gamma=0.15\pm 0.03$ for the present dark energy density parameter $\Omega_{\Lambda 0}=0.73\pm0.03$. 

\end{abstract}


\section{Introduction}
 Recent observations of type Ia supernovae (SNIa) have revealed that 
 the present Universe is undergoing accelerated expansion~\cite{riess1}.
This indicates that the energy density of the Universe at present epoch is dominated by dark energy with equation of state $w_{\Lambda} = \frac{p_{\Lambda}}{\rho_{\Lambda}}<-\frac{1}{3}$, 
where $\rho_{\Lambda}$ is the energy density of dark energy, 
and $p_{\Lambda}$ is its pressure.
Combining with observations of CMB anisotropy~\cite{WMAP} 
 and the baryon acoustic oscillation (BAO)~\cite{SDSS}, 
 the density parameter is determined as 
 $\Omega_{\Lambda 0} = \rho_{\Lambda 0}/\rho_{c 0} \sim 0.7$, 
 where $\rho_{c 0} = 8.0992h^{2}\times 10^{-47}{\rm GeV}^{4}$ is the critical density today 
 and $H_{0}=100h~{\rm km/s/ Mpc}$ is the present Hubble parameter.
 
 The simplest way to explain dark energy is to introduce 
 a cosmological constant~\cite{einstein}. 
 However, this model suffers from 
 two problems~\cite{weinberg}.
 The first one is the fine-tuning problem: 
the observed cosmological constant is extremely small compared to the
fundamental Planck scale $\rho_{\Lambda}$ $\sim$ $10^{-120} M_{p}^{4}$, 
requiring an incredible fine-tuning. The second one is the cosmic 
coincidence problem: why the cosmological constant and matter 
have comparable energy density today even though their time evolution 
is so different.
 A variety of models have been proposed to solve these problems, 
 such as quintessence~\cite{quin}, phantom~\cite{phantom}, 
 quintom~\cite{quintom} 
 and holographic dark energy~\cite{li, gao, cosmicconstraintsRDE, RDE}. 
 
 In particular, the holographic dark energy (HDE) models 
 have been discussed extensively in recent years.
 These models are 
 motivated by 
 the holographic
 principle of quantum gravity~\cite{quantumgravity}.
 From the condition that a system with size $L$ 
 would not form a black hole, 
 it is required that
 the total vacuum energy should not 
 exceed the mass of the black hole of the 
 same size. 
 Therefore, the dark energy 
 density 
 $\rho_{\Lambda}$ must satisfy 
 $L^{3} \rho_{\Lambda}\lesssim L M_{p}^{2}$~\cite{cohen}, 
 where $M_{p} = 1/\sqrt{8\pi G}$ is the reduced Planck mass. 
 Saturating this inequality, 
 $\rho_{\Lambda}$ is defined by~\cite{li}
  \begin{eqnarray}
   \rho_{\Lambda}=3c^{2}M_{p}^{2}L^{-2},
  \end{eqnarray}
 where $c$ is a dimensionless parameter. 
 In Ref.~\cite{cohen}, the size $L$, which is regarded as 
 an IR cut-off, was chosen to be 
 the inverse Hubble expansion rate $L\sim H^{-1}$ 
 to naturally explain the observed vacuum energy density 
 $\rho_{\Lambda 0} \sim 3 M_{p}^{2}H_{0}^{2}$. 
 However, it was shown that this choice can not explain the 
 accelerated expansion of the Universe at present~\cite{hsu}. 
 This problem was solved in Ref.~\cite{li} by choosing $L$ 
 to be the future event horizon $R_{h}$. 
 Extensive studies on this 
 model have been done, and the HDE model with the future event horizon 
 as the IR cut-off is found to be 
 consistent with current observational data~\cite{consistentobs}. 
 Various aspects on the HDE model have been discussed in Ref.~\cite{aspects}. 
 There have also been further developments on the HDE model by 
introducing an interaction 
between dark energy and matter~\cite{IHDE, newinteraction}. 
 However, it was pointed out that this model 
 with $L=R_{h}$ has a conceptual problem that 
 the future event horizon, which determines $\rho_{\Lambda}$, 
 depends on the future evolution of the Universe, 
 hence violates causality~\cite{causality}.

Subsequently, 
inspired by the HDE model, 
the holographic Ricci dark energy (RDE) model~\cite{gao} 
was proposed in which $\rho_{\Lambda}$ is proportional to the 
Ricci scalar curvature ${\cal R}$. It was shown that this model does not 
only avoid the causality problem and is phenomenologically viable, 
but also naturally solves the coincidence problem. 
Also, it was found that the causal connection scale 
$R_{\rm CC}$ consistent with cosmological observations is 
given by $R_{\rm CC}^{-2}=\dot{H}+2 H^{2}$ 
which is proportional to ${\cal R}$ in a flat universe~\cite{Rcc}. 
This may provide us with a physical motivation for the RDE model.
Cosmological constraints on this model 
were studied in Ref.~\cite{cosmicconstraintsRDE}. 
Similar models have also been studied in Ref.~\cite{RDE}.

In this paper, we consider the RDE model with an interaction between 
dark energy and matter, 
and call it 
the interacting Ricci dark energy (IRDE) model. 
We investigate the observational constraints on this model 
obtained from SNIa, CMB and BAO data. 
We organize this paper as follows. 
In section 2, we describe the IRDE model, 
and obtain
analytic expressions for cosmic time evolution. 
In section 3, 
we discuss the observational constraints on 
this model. 
We summarize our results in section 4.


\section{The interacting Ricci Dark Energy model}
We consider the spatially 
homogeneous
and isotropic 
Universe described by 
the Friedmann-Robertson-Walker metric 
\begin{eqnarray}
 ds^{2}= dt^{2} - a^{2}(t)
                \left(
                         \frac{dr^{2}}{1-kr^{2}} +r^{2}d \theta^{2} 
                                   +r^{2}\sin^{2}{\theta} d \phi^{2}
                \right),
 \end{eqnarray}
where $k=1, 0, -1$ for closed, flat and open geometries.
The time evolution of the scale factor $a(t)$ is described by the Friedmann equation
 \begin{eqnarray}
  H^{2} = \frac{1}{3M_{p}^{2}} 
                (\rho_{\Lambda}+\rho_{m}+\rho_{\gamma}+\rho_{k}),
  \label{Friedmann}
  \end{eqnarray}
where $\rho_{\Lambda}, \rho_{m}, \rho_{\gamma}$ and $\rho_{k}$ represent energy density of dark energy, 
matter, radiation and curvature, respectively, and $H=\dot{a}/a$ is the Hubble parameter.

The energy density of dark energy in the IRDE model is defined as~\cite{gao}
 \begin{eqnarray}
 \label{rde}
  \rho_{\Lambda}
       &=& 3\alpha M_{p}^{2}
       \left (
       \dot{H}+2H^{2} + \frac{k}{a^{2}}
       \right),
                \label{RDE}
 \end{eqnarray}
 where $\alpha$ is a dimensionless parameter.
 Note that $\rho_{\Lambda}$ is proportional to the Ricci scalar curvature
 \begin{eqnarray}
 {\cal{R}}= -6
                      \left(\dot{H}+2H^{2}+\frac{k}{a^{2}}
                      \right).
 \end{eqnarray}
 Moreover, it is assumed that there is an interaction 
 between dark energy and matter. 
The energy densities $\rho_{\Lambda}$ and $\rho_{m}$ obey the following equations~\cite{IHDE}
 \begin{eqnarray}
 \label{CLL}
  \dot{\rho_{\Lambda}} + 3H (1+w_{\Lambda})\rho_{\Lambda} &=& -Q,\\
  \label{CLM}
    \dot{\rho_{m}} + 3H \rho_{m} &=& Q.
   \end{eqnarray}
The interaction rate is given by
\footnote{
If the interaction rate is given by 
$Q=9b^{2}M_{p}^{2}H^{3}~(b^{2}>0)$~\cite{IHDE}, 
it turns out that the matter density $\rho_{m}$ becomes negative for $a \ll 1$. 
This problem does not occur for $b^{2} < 0$. 
However, the case $b^{2}<0$ 
is disfavored by observational constraints. }~\cite{newinteraction} 
 \begin{eqnarray}
 \label{intrate}
   Q=\gamma H \rho_{\Lambda},
  \end{eqnarray}
  where $\gamma$ is a dimensionless parameter.
The energy density of radiation is given by 
$\rho_{\gamma}=\rho_{\gamma 0}a^{-4}$, where $\rho_{\gamma 0}$ 
is the present value of radiation density.
We adopt a convention that $a(t_{0})=1$ for the present age of the Universe 
$t_{0} \approx 14$ Gyr. 
According to eq.~(\ref{CLM}) with $Q$ in eq.~(\ref{intrate}), 
the interaction can be relevant if $\gamma \rho_{\Lambda}$ and $\rho_{m}$ are 
comparable, whether or not the Universe is in the radiation-dominated epoch. 
In literatures, various types of interactions between dark energy and 
matter have been considered, including $Q$ = $\gamma M_P^2 H^3$, 
$\gamma H \rho_{\Lambda}$, $\gamma H \rho_{m}$, 
$\gamma H (\rho_{\Lambda}+\rho_{m})$, and so forth. At present, no definite 
mechanisms to determine the interaction are established. Thus, we simply assume 
eq. (\ref{intrate}) as a phenomenological model. 

 Combining with eqs. (\ref{RDE}) and (\ref{CLM}), the Friedmann equation (\ref{Friedmann})
 is written as 
  \begin{eqnarray}
\nonumber
    \frac{ \alpha}{2} \frac{d^{2}H^{2}}{dx^{2}} 
    - \left(1-\frac{7\alpha}{2} -\frac{\alpha \gamma}{2} \right)\frac{d H^{2}}{dx} 
    -(3-6\alpha-2\alpha \gamma) H^{2}  
    &&\\
\label{EOH}    
               - \frac{\rho_{\gamma 0}}{3M_{p}^{2}} e^{-4x}
                -\{1- \alpha(1+\gamma ) \}  k e^{-2x}
  &=&0,
 \end{eqnarray}
 where $x=\ln{a}$. 
The solution to eq. (\ref{EOH}) is obtained as 
 \begin{eqnarray}
 \label{hubbleparameter}
  \frac{H^{2}}{H_{0}^{2}}    &=& A_{+} e^{\sigma_{+} x} + A_{-} e^{\sigma_{-} x}
            + A_{\gamma} e^{-4 x} + A_{k} e^{-2 x} ,
\label{solution}
 \end{eqnarray}
where 
 \begin{eqnarray}
 \label{sigmapm}
  \sigma_{\pm} 
   &=& \frac{2-7\alpha-\alpha \gamma \pm\sqrt{(2-\alpha)^{2}
   -2\alpha (\alpha +2) \gamma+\alpha^{2}\gamma^{2}}}{2\alpha},
 \end{eqnarray}
   $\Omega_{\gamma 0}=\rho_{\gamma 0}/\rho_{c0}$,  
 $\Omega_{k 0}= - k/H_{0}^{2}$ and  
 $\rho_{c0}=3M_{p}^{2}H_{0}^{2}$. 
 Note that $\sigma_{\pm}$ can be imaginary for 
 sufficiently large $\alpha$ and $\gamma$. 
 This implies that there is a parameter region where $H^{2}$
 has oscillatory behavior. 
 However, this region is not phenomenologically viable.
 The constants $\Omega_{\gamma 0}$ and $\Omega_{k 0}$ are the present 
 value of $\Omega_{\gamma}$ and $\Omega_{k}$, respectively.
 The constants $A_{\gamma}$, $A_{k}$ and $A_{\pm}$ are given by
  \begin{eqnarray}
  A_{\gamma} &=&  \Omega_{\gamma 0} ,
 \end{eqnarray}
 \begin{eqnarray}
  A_{k} &=&   \Omega_{k 0} ,
 \end{eqnarray}
 \begin{eqnarray}
 \label{rholambda}
 A_{\pm} &=& \pm \frac{ 
            \alpha(\sigma_{\mp}+3) \Omega_{k 0}
                + 2 \Omega_{\Lambda 0}
                -\alpha (1-\Omega_{\gamma 0})(\sigma_{\mp}+4)  }
               {\alpha(\sigma_{+}-\sigma_{-})}.  
 \end{eqnarray}
Without the interaction ($\gamma=0$), eq. (\ref{solution}) reduces   to the result obtained 
in the literature~\cite{gao}.
In this case, the constants in eq. (\ref{sigmapm}) 
are $\sigma_{+}=-4 +2/\alpha$ and $\sigma_{-}=-3$.

 Substituting eq. (\ref{solution}) to eq. (\ref{RDE}), 
 the Ricci dark energy density is given by
 \begin{eqnarray}
 \label{rdes}
  \rho_{\Lambda}
    &=&   \rho_{c0}  
              \sum_{i=+,-} \alpha
              \left(\frac{ \sigma_{i}}{2}+2
                                    \right)
                                    A_{i} e^{\sigma_{i}x} .
 \end{eqnarray}
 Likewise, the matter density is 
 \begin{eqnarray}
 \label{rms}
  \rho_{m}
    &=&  \rho_{c0}
              \sum_{i=+,-}
             \left\{1-
                       \alpha \left(\frac{ \sigma_{i}}{2} +2 \right)
             \right\}
                         A_{i} e^{\sigma_{i}x}.
  \end{eqnarray}
The equation of state of dark energy 
 can be found by 
 substituting eq. (\ref{rdes}) into the following expression:
 \begin{eqnarray}
  w_{\Lambda} 
    =  -1-\frac{1}{3}
             \left(
                      \gamma + \frac{1}{\rho_{\Lambda}}\frac{d\rho_{\Lambda}}{dx}
             \right).
  \end{eqnarray}

In eqs.~(\ref{rdes}) and (\ref{rms}), 
the term proportional to  $e^{\sigma_{-}x}$ is dominant in the past $a$ $\ll$ 1, while 
the term proportional to $e^{\sigma_{+}x}$ is dominant in the future $a$ $\gg$ 1. 
As an illustration, let us consider the case $\alpha$ $=$ 0.45 and $\gamma$ 
$=$ 0.15 which corresponds to $\sigma_{+}$ $\approx$ 0.25 and 
$\sigma_{-}$ $\approx$ $-3.0$. 
In the past $a$ $\ll$ 1, the ratio of eq.~(\ref{rms}) to eq.~(\ref{rdes}) is 
$\rho_m/\rho_{\Lambda}$ $\approx$ $\alpha^{-1}(2+\sigma_{-}/2)^{-1}-1$ $\approx$ 3.4, 
while $\rho_m/\rho_{\Lambda}$ $\approx$ $\alpha^{-1}(2+\sigma_{+}/2)^{-1}-1$ 
$\approx$ 0.045 in the future $a$ $\gg$ 1.

 Notice that both $\rho_{\Lambda}$ and $\rho_{m}$ include contributions 
 with non-standard time evolution. 
 This typically leads to a constant ratio of $\rho_{\Lambda}$ to $\rho_{m}$, 
 and it may help to solve the coincidence problem. 
 As pointed out in Ref.~\cite{gao}, the coincidence problem is ameliorated 
in the original RDE model without interaction where 
$\rho_{\Lambda}$ and $\rho_{m}$ were comparable with each other in the 
past universe. In this case, $\rho_{\Lambda}$ starts to increase at low redshift, 
and the ratio $\rho_{\Lambda}/\rho_{m}$ rapidly grows in the future, since 
$\rho_{m} \sim e^{-3x}$ in the absence of interaction. On the other hand, in the 
IRDE model, the behavior in the past is similar to that in the original 
RDE model, but the ratio $\rho_{\Lambda}/\rho_{m}$ is constant even in 
the future. 

\section{ Observational constraints}
In this section, we study the cosmological constraints on the 
IRDE 
model obtained from SNIa, CMB and BAO observations. 
In what follows, we focus only on the spatially flat Universe ($k=0$).

The SNIa observations measure the distance modulus $\mu$ 
of a supernova and its redshift $z$. 
The distance modulus is defined by
 \begin{eqnarray}
  \mu = 5 \log_{10}{\frac{d_{L}}{\rm Mpc}},
 \end{eqnarray}
 where $d_{L}(z)$ is the luminosity distance given by
  \begin{eqnarray}
 d_{L} = (1+z) \int_{0}^{z} \frac{dz'}{H(z')}.
 \end{eqnarray}
 We apply the Union data set of 307 SNIa~\cite{307SN}, 
 and place limits on the parameters $\alpha, \gamma$ and $\Omega_{\Lambda 0}$ 
 by minimizing~\cite{chi2}
 \begin{eqnarray}
  \chi_{\rm SN}^{2} 
   = A-\frac{B^{2}}{C},
 \end{eqnarray}
 where 
 \begin{eqnarray}
  A &=& \sum_{i=1} ^{307}
                 \left(
                         \frac{\mu(z_{i})-\mu_{\rm obs}(z_{i})}{\sigma_{\mu_{i}}}
                 \right)^{2},\\
   B &=&  \sum_{i=1} ^{307}
                         \frac{\mu(z_{i})-\mu_{\rm obs}(z_{i})}{\sigma_{\mu_{i}}^{2}},\\
   C &=& \sum_{i=1}^{307} \frac{1}{\sigma_{\mu_{i}}^{2}}.
 \end{eqnarray}
 Here, $z_{i}$ and $\mu_{\rm obs}(z_{i})$ are the redshift and the distance modulus of the $i$-th supernova,
 respectively.
 The corresponding $1\sigma$ error is denoted by $\sigma_{\mu_{i}}$.

The CMB shift parameter is one of the least model dependent parameters extracted from the CMB data.
Since this parameter involves the large redshift behavior ($z\sim 1000$), 
it gives a complementary bound to the SNIa data ($z\lesssim 2$).
The shift parameter $R$ is defined as 
 \begin{eqnarray}
  R &=& \sqrt{\Omega_{m0}} 
                 \int_{0}^{z_{\rm CMB}} \frac{H_{0}}{H(z)} dz, 
 \end{eqnarray}
 where
 $z_{\rm CMB}=1090$ is the redshift at recombination, 
 and $\Omega_{m0}=\rho_{m0}/\rho_{c0}$ is the matter fraction at present. 
 We use the value $R=1.710\pm0.019$ obtained 
 from the WMAP5 data~\cite{WMAP5}. 
 The CMB constraints are given by minimizing
  \begin{eqnarray}
  \chi_{\rm CMB}^{2} 
   =  
                 \left(
                         \frac{R-R_{\rm obs}}{\sigma_{R}}
                 \right)^{2},
 \end{eqnarray} 
 where  $R_{\rm obs}=1.710$ and $\sigma_{R}=0.019$. 
 
Observations of large-scale galaxy clustering provide the signatures of
the baryon acoustic oscillation (BAO). 
We use the measurement of the BAO peak in the distribution of luminous red galaxies (LRGs) observed 
 in SDSS~\cite{SDSS}. 
 It gives
 \begin{eqnarray}
  A &=& 0.469
                         \left(
                                  \frac{0.95}{0.98}
                         \right)^{-0.35}
               \pm 0.017,
 \end{eqnarray}
 where the parameter $A$ is given by
  \begin{eqnarray}
  A &=& \sqrt{\Omega_{m0}}
               \left(
                       \frac{H_{0}}{H(z_{\rm BAO})}
               \right)^{1/3}
               \left[
                       \frac{1}{z_{\rm BAO}} \int_{0}^{z_{\rm BAO}}  \frac{H_{0}}{H(z)}dz
               \right]^{2/3},
     \end{eqnarray}
and $z_{\rm BAO}=0.35$. 
The $\chi^{2}$ for BAO is
 \begin{eqnarray}
  \chi_{\rm BAO}^{2} 
   = 
                 \left(
                         \frac{A-A_{\rm obs}}{\sigma_{A}}
                 \right)^{2},
 \end{eqnarray}
   where $A_{\rm obs}=0.469(0.95/0.98)^{-0.35}$ and 
   $\sigma_{A}=0.017$ .\\
   
In Fig.1, we plot the probability contours for SNIa (red),
CMB (blue), BAO (green) observations in the ($\alpha, \Omega_{\Lambda 0}$)-plane in the case without interaction ($\gamma=0$).
The $1\sigma$, $2\sigma$ and $3\sigma$ contours are drawn with
solid, dashed and dotted lines, respectively.
The joined constraints using $\chi^{2}=\chi_{\rm SN}^{2}+\chi_{\rm CMB}^{2}+\chi_{\rm BAO}^{2}$ are shown as shaded contours. 
The best fit values with $1\sigma$ error are $\alpha = 0.36 \pm 0.03$ 
and $\Omega_{\Lambda 0}= 0.68\pm0.03$
with $\chi_{\rm min}^{2}=326$.
 The joint $1\sigma$ region is outside the $1\sigma$ regions of CMB and BAO.
 
\begin{figure}[htbp]
\centering
\includegraphics[width=3.8in]{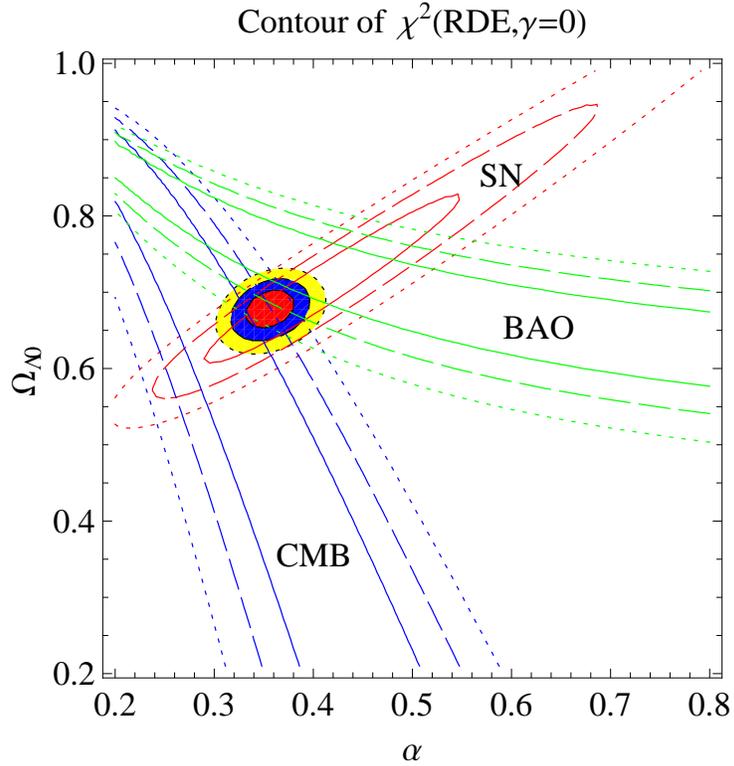}
\caption{
The probability contours for SNIa (red),
CMB (blue), BAO (green) observations in the ($\alpha, \Omega_{\Lambda 0}$)-plane in the case without interactions ($\gamma=0$). 
The $1\sigma$, $2\sigma$ and $3\sigma$ contours are drawn with
solid, dashed and dotted lines, respectively.
The joined constraints using 
$\chi^{2}=\chi_{\rm SN}^{2}+\chi_{\rm CMB}^{2}+\chi_{\rm BAO}^{2}$ 
are shown as shaded contours. 
The flat Universe ($k=0$) is assumed in all the figures.
}\label{fig:b=0}
\end{figure}

\begin{figure}[htbp]
\centering
\includegraphics[width=3.8in]{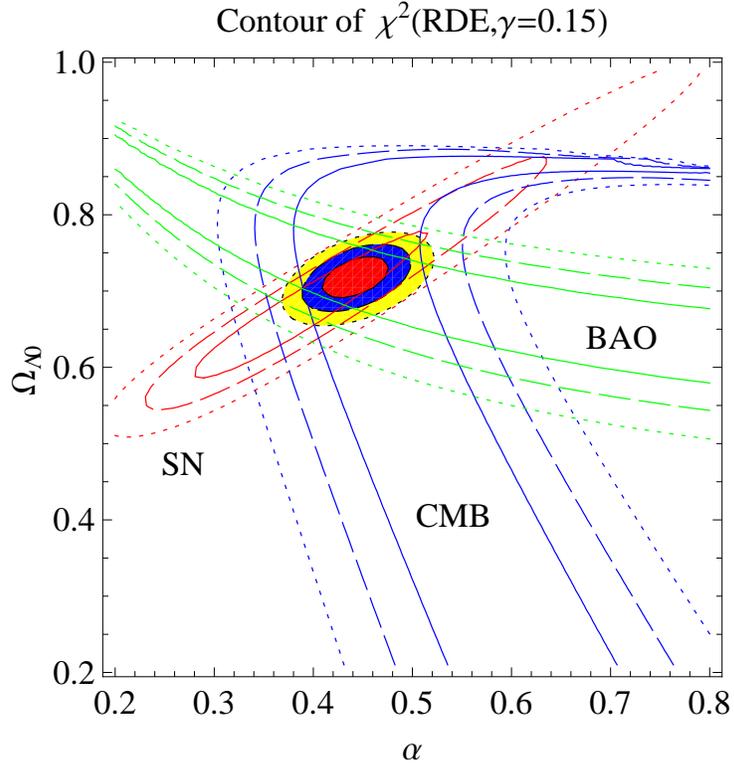}
\caption{
The same contour plots as Fig.\ref{fig:b=0} 
 in the presence of interaction with $\gamma=0.15$.
  }\label{fig:b=0.13}
\end{figure}

 The same contour plots as Fig.\ref{fig:b=0} 
 in the presence of interaction with $\gamma=0.15$ 
 are shown in Fig.\ref{fig:b=0.13}. 
 Compared to Fig.1, the CMB contours are shifted to the right.
 The best fit values are shifted as
 $\alpha=0.45$ and $\Omega_{\Lambda 0}=0.73$. 
 It is remarkable to note that, 
 unlike the case $\gamma=0$, the joined $1\sigma$ region is included
 in all the $1\sigma$ regions of SNIa, CMB and BAO.
 
 Dependence on the interaction parameter $\gamma$ is presented in the 
 ($\gamma, \alpha$)-plane for $\Omega_{\Lambda 0}=0.73$
 in Fig.3.
 The best fit values are $\gamma=0.15 \pm0.03$ and $\alpha=0.45\pm0.03$ 
 with  $\sigma_{+}=0.25$, $\sigma_{-}=-3.0$ and 
 $\chi_{\rm min}^{2}=313$.
 It is found that the CMB constraint strongly depends on $\gamma$.
 As a result, the joined $1\sigma$ region appears in the range with sizable 
 $\gamma$.
 This implies that existence of interaction between 
 dark energy and matter is favored 
 by cosmological constraints in the IRDE model.

\begin{figure}[htbp]
\centering
\includegraphics[width=3.8in]{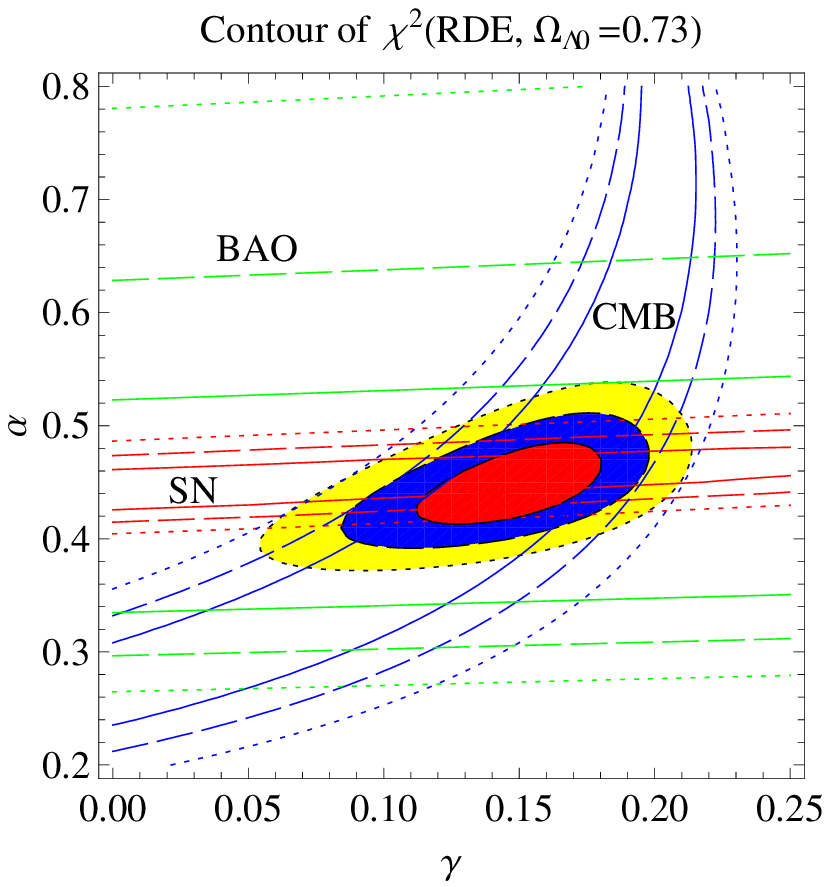}
\caption{
The contours of $\chi^{2}$ in the ($\gamma, \alpha$)-plane.
}
\label{fig:alpha_b}
\end{figure}
 \section{Conclusions}
 We have considered the IRDE model 
 where the interaction rate is given by eq. (\ref{intrate}). 
 We have derived the analytic expressions 
 for the Hubble parameter (\ref{hubbleparameter}) 
 and the energy density of dark energy (\ref{rdes}) and matter (\ref{rms}).  
 Both $\rho_{\Lambda}$ and $\rho_{m}$ include contributions 
 with non-standard time evolution.
 We have also investigated current observational constraints on this model 
 from SNIa, CMB and BAO observations. 
 In particular, the CMB constraint is strongly affected by the interaction.
 We have shown that a nonvanishing interaction rate is favored by 
 the observations, 
 giving a reduction of $\chi_{\rm min}^{2}$. 
 The best fit values are $\Omega_{\Lambda 0}=0.73\pm0.03$, $\alpha=0.45\pm0.03$ and $\gamma=0.15\pm0.03$.


  \end{document}